\documentstyle[11pt,equation]{article}

\font\tenrm=cmr10
\font\tenit=cmti10
\font\elevenbf=cmbx10 scaled\magstep 1
\font\elevenrm=cmr10 scaled\magstep 1
 1

\input{margin}
\renewenvironment{thebibliography}[1]
 { \elevenrm
   \begin{list}{\arabic{enumi}.}
    {\usecounter{enumi} \setlength{\parsep}{0pt}
     \setlength{\itemsep}{3pt} \settowidth{\labelwidth}{#1.}
     \sloppy
    }}{\end{list}}

\newcommand{\roots}{\mbox{$\sqrt{s}$}}
\newcommand{\ccbar}{\mbox{$c \overline{c} $}}
\newcommand{\bbbar}{\mbox{$b \overline{b} $}}
\newcommand{\ttbar}{\mbox{$t \overline{t} $}}
\newcommand{\Y}{\mbox{$\Upsilon$}}

\newcommand{\shat}{\mbox{$\hat{s}$}}
\newcommand{\lhat}{\mbox{$\hat{\cal L}$}}

\newcommand{\rs}{\mbox{$\sqrt{s}$}}

\newcommand{\mjj}{\mbox{$M_{jj}$}}

\newcommand{\ee}{\mbox{$e^+e^-$}}
\newcommand{\gaga}{\mbox{$\gamma\gamma$}}
\newcommand{\be}{\begin{equation}}
\newcommand{\ene}{\end{equation}}
\newcommand{\een}{\end{subequations}}
\newcommand{\ben}{\begin{subequations}}
\newcommand{\beq}{\begin{eqalignno}}
\newcommand{\eeq}{\end{eqalignno}}

\begin{document}
\begin{flushright}
MAD/PH/776 \\
BU--TH/93--3 \\
\end{flushright}
\vspace*{2cm}
\begin{center}
\vglue 0.6cm
{{\elevenbf HARD \gaga\ COLLISIONS AT FUTURE \ee\ LINACS\footnote{To appear
in the proceedings of the Linear Collider Workshop, Waikoloa, Hawaii, April
1993.}\\}
\vglue 1.0cm
{\tenrm Manuel Drees\footnote{Heisenberg fellow} \\}
\baselineskip=13pt
{\tenit Physics Department, University of Wisconsin, Madison, Wi53706, USA\\}

\vglue 0.3cm
{\tenrm and\\}
\vglue 0.3cm
{\tenrm Rohini M. Godbole\\}
{\tenit Physics Department, University of Bombay, Bombay 400098, India\\}}
\vglue 0.8cm
{\tenrm ABSTRACT}

\end{center}

\vglue 0.3cm
{\rightskip=3pc
 \leftskip=3pc
 \tenrm\baselineskip=12pt
 \noindent
We update our previous calculation of di--jet production in two--photon
collisions at future \ee\ colliders, including both beam-- and
bremsstrahlung. We then discuss the production of heavy ($c,b,t$) quarks. Our
study re--confirms that colliders with much beamstrahlung offer no physics
advantages over designs with little beamstrahlung.}
\setcounter{page}{0}
\clearpage
\pagestyle{plain}
\baselineskip=14pt
\elevenrm
\vglue 0.6cm
{\elevenbf\noindent 1. Introduction}
\vglue 0.4cm
It is by now well known \cite{1} that a substantial contribution to the
(effective) photon flux at future high--energy \ee\ colliders (linacs) will
come from beamstrahlung. In this contribution we discuss the impact of
beamstrahlung on the production of jet pairs with invariant mass well
below the nominal \rs\ of the machine (sec. 2), as well as on the production
of heavy quarks (sec. 3).
\baselineskip=12pt
\tenrm
\vglue 0.5cm
\noindent
{\bf Table 1:} Parameters of the machine designs we use, from
ref.\cite{2}. \rs\ is the nominal center--of--mass energy, \Y\ the
beamstrahlung parameter, $\sigma_z$ the bunch length, \lhat\ the luminosity
per bunch crossing, and ${\cal L}$ the luminosity of the collider. Notice that
beam disruption effects have been included in \Y, \lhat\ and ${\cal L}$.
\begin{center}
\begin{tabular}{|c||c|c|c|c|c|}
\hline
Collider & \roots\ [TeV] & \Y\ & $\sigma_z$ [mm]& \lhat\ [$\mu b^{-1}$]  &
${\cal L} [10^{33}/$cm$^2$ sec] \\
\hline
CLIC & 0.5 & 0.35 & 0.17 & 1.3 & 8.8 \\
DLC & 0.5 & 0.071 & 0.5 & 0.76 & 6.67 \\
JLC & 0.5 & 0.15 & 0.08 & 0.74 & 10.1 \\
NLC & 0.5 & 0.096 & 0.1 & 0.51 & 8.22 \\
TESLA & 0.5 & 0.065 & 1.0 & 1.39 & 11.1 \\
\hline
DLC & 1.0 & 0.42 & 0.17 & 3.7 & 9.2 \\
JLC & 1.0 & 0.38 & 0.113 & 3.1 & 12.8 \\
NLC & 1.0 & 0.27 & 0.1 & 2.18 & 17.5 \\
TESLA & 1.0 & 0.24 & 1.1 & 5.86 & 46.6 \\
\hline
\end{tabular}
\end{center}
\baselineskip=14pt
\elevenrm
\vglue 0.5cm
\noindent
Since beamstrahlung depends on many quantities describing the charge and size
of bunches, we first have to fix the parameters of the colliders that we are
going to study, see table 1; these have been taken from ref.\cite{2}, and
roughly describe the ``state of the art'' of collider design as of fall 1992.
{}From these parameters we compute the photon and electron beam spectra using
Chen's approximate expressions \cite{3}, which should be applicable to
these designs. The resulting effective single photon beamstrahlung spectra
$f^{\rm beam}_{\gamma|e}$, averaged over a bunch collision, are shown in figs.
1 a,b. We see that, with the exception of CLIC, beamstrahlung dominates over
bremsstrahlung only for relatively small photon energies, $x \leq 0.2$, at
\rs\ = 500 GeV; however, at \rs\ = 1 TeV the cross--over occurs at much
larger values of $x$, between 0.4 and 0.7. For all designs beamstrahlung
enhances the flux of soft photons significantly (NLC, JLC) or even
dramatically (DLC, TESLA); it should be mentioned that the versions of the
last two designs given in table 1 actually produce considerably more
beamstrahlung than older versions \cite{4}.
\vglue 0.6cm
{\elevenbf\noindent 2. Di--Jet Production}
\vglue 0.4cm
The formalism for calculating jet cross sections in the presence of
beamstrahlung has been described in some detail in ref.\cite{5}. Here we
merely remind the reader that three classes of processes contribute to
the reaction $\gaga \rightarrow $jets, depending on whether the initial state
photons interact directly or via their hadronic constituents \cite{6}.
Following the notation of ref.\cite{7} we call these contributions direct,
1--resolved and 2--resolved, respectively. The existence of resolved photon
contributions is by now firmly established experimentally \cite{8}; however,
the parton densities inside the photon at small Bjorken$-x$, as well as
the gluon density at all $x$, are at present still only weakly constrained. We
try to estimate the resulting uncertainty by using several parametrizations
\cite{dg,grv,lac} for these distribution functions. As discussed in more
detail in ref.\cite{9}, analyses of TRISTAN and HERA data should substantially
reduce this uncertainty in the near future.

Ref.\cite{5} contains a fairly exhaustive discussion of jet production at
\ee\ linacs. For reasons of space we here only update our calculation of the
invariant mass spectrum of central di--jet pairs well below the nominal \rs\
of the collider. There are two motivations for studying annihilation events
with \mjj $<\rs$. Firstly, one might be able to study the QCD evolution of
the jet system with increasing \mjj\ with a single detector, thereby hopefully
reducing experimental uncertainties. Secondly, the peak at $\mjj=M_Z$ might
offer an opportunity to precisely calibrate the detector. Clearly neither of
these tasks can be performed if most events with $\mjj \ll \rs$ actually
come from \gaga\ processes.

In fig. 2 we therefore compare \gaga\ and annihilation contributions for three
representative designs of 500 GeV colliders. Both beamstrahlung and
bremsstrahlung or initial state radiation have been included. We have required
$p_T({\rm jet}) > 20$ GeV, which reduces the two--photon background
considerably but keeps most of the annihilation events. Nevertheless we see
that at CLIC, which has the by far hardest beamstrahlung spectrum (see fig.
1a), the two--photon contribution becomes negligible only for $\mjj \geq 300$
GeV. In particular, there will be no pronounced peak at $\mjj = M_Z$ at this
collider. Reducing beamstrahlung reduces both signal and background; it is
important to keep in mind that one cannot produce a substantial flux of hard
photons without simultaneously producing a sizable flux of electrons with
energy $< \rs/2$. Nevertheless the signal--to--background ratio is greatly
improved at machines with less beamstrahlung, as shown by the solid and
short--dashed curves; in particular, at these colliders a clear $Z$ peak
should be visible, the integrated signal being about 1 pb or several $10^4$
events per year. However, for all designs of table 1 the region just above the
$Z$ peak is again dominated by the two--photon contribution; this is to be
contrasted with the 1991 TESLA design, see fig. 7b of ref.\cite{5}. We finally
mention that, due to the $p_T$ cut, the two--photon contribution in fig. 2
comes mostly from direct processes; even at CLIC the structure function
uncertainty in the total two--photon contribution only amounts to 5\%.
\vglue 0.6cm
{\elevenbf\noindent 3. Heavy Quark Production}
\vglue 0.4cm
Previous studies have found large rates for the two--photon production of
$c$ and $b$ \cite{5,10} and even $t$ quarks \cite{10}. In tables 2 and 3 we
present total \ccbar\ and \bbbar\ cross sections for the colliders listed in
table 1, as well as for an ideal machine without beamstrahlung (\Y\ = 0). As
in refs.\cite{5,10} we find very large rates, mostly due to beamstrahlung.
Notice also that resolved photon contributions are very important.
Unfortunately they are also very uncertain, since we are probing the gluon
density at small $x$. At present we can therefore estimate the total rates
only up to a factor 3 to 5 for \ccbar\ and 2 to 3 for \bbbar\ production;
recall, however, that the same $x-$range can be probed in photoproduction at
HERA, so that the large rates predicted by the LAC1 parametrization should be
either confirmed or excluded experimentally in the near future, thereby
reducing the uncertainty of the present calculation.

In spite of this uncertainty we can conclude from table 2 that already at
\rs\ = 500 GeV the total \ccbar\ cross section from two--photon processes
will be of similar magnitude as or even larger than the open \ccbar\
cross section at a $\tau/$charm factory ($\simeq 10$ nb) or on the $Z$
pole at LEP1 ($\simeq 5$ nb). However, the ratio of hadronic non--charm to
charm events is considerably higher in \gaga\ collisions than in \ee\
annihilation. Even the rather conservative ``reference model'' of ref.\cite{2}
predicts the total hadronic rates to be some 20 times higher than the DG
prediction for the total \ccbar\ rate. Some sort of charm tagging will
therefore be necessary for the study of charmed events in \gaga\ collisions;
of course, the situation is even worse for \bbbar\ production.
\vglue 0.5cm
\tenrm
\baselineskip=12pt
\noindent
{\bf Table 2:} Total \ccbar\ cross sections from two--photon processes at the
\ee\ colliders of table 1; for comparison we also give results for negligible
beamstrahlung (\Y=0). The first 6 rows are for \rs= 0.5 TeV, and the
remaining entries are for \rs=1 TeV. The subscripts ``DG'', ``GRV'' and
``LAC1'' refers to the parametrizations of refs.\cite{dg},\cite{grv} and
\cite{lac}, respectively, while the superscripts ``dir'', ``1--res'' and
``2--res'' stand for the direct, 1--resolved and 2--resolved
contributions to the cross section. The resolved photon contributions have
been computed using $Q^2 = \shat/4$ and a $Q^2$--dependent number of active
flavors \cite{5}. All cross sections are in nb.
\begin{center}
\begin{tabular}{|c||c||c|c||c|c||c|c|}
\hline
Collider & $\sigma^{\rm dir}$ & $\sigma_{\rm DG}^{\rm 1-res}$ &
$\sigma_{\rm DG}^{\rm 2-res}$ & $\sigma_{\rm GRV}^{\rm 1-res}$ &
$\sigma_{\rm GRV}^{\rm 2-res}$ & $\sigma_{\rm LAC1}^{\rm 1-res}$ &
$\sigma_{\rm LAC1}^{\rm 2-res}$ \\
\hline
CLIC & 15 & 51 & 2.3 & 31 & 1.8 & 300 & 29 \\
DLC & 9.6 & 11 & 0.35 & 7.6 & 0.28 & 57 & 2.7 \\
JLC & 2.2 & 4.2 & 0.16 & 2.7 & 0.13 & 24 & 1.9 \\
NLC & 1.8 & 2.9 & 0.10 & 1.9 & 0.084 & 16 & 1.2 \\
TESLA & 26 & 25 & 0.75 & 18 & 0.62 & 120 & 4.7 \\
\Y=0 & 0.55 & 0.91 & 0.033 & 0.54 & 0.026 & 5.2 & 0.53 \\
\hline
DLC & 38 & 300 & 19 & 160 & 15 & 1,800 & 360 \\
JLC & 3.5 & 23 & 1.4 & 12 & 1.1 & 130 & 29\\
NLC & 2.4 & 12.2 & 0.7 & 6.6 & 0.55 & 71 & 14 \\
TESLA & 76 & 410 & 23 & 240 & 18 & 2,400 & 360 \\
\Y=0 & 0.72 & 2.0 & 0.094 & 1.1 & 0.072 & 11 & 1.9 \\
\hline
\end{tabular}
\end{center}
\vspace*{6mm}
\noindent{\bf Table 3:} Total cross sections for \bbbar\ production from
two--photon processes. The notation is as in table 2, except that now all
cross sections are in pb.
\begin{center}
\begin{tabular}{|c||c||c|c||c|c||c|c|}
\hline
Collider & $\sigma^{\rm dir}$ & $\sigma_{\rm DG}^{\rm 1-res}$ &
$\sigma_{\rm DG}^{\rm 2-res}$ & $\sigma_{\rm GRV}^{\rm 1-res}$ &
$\sigma_{\rm GRV}^{\rm 2-res}$ & $\sigma_{\rm LAC1}^{\rm 1-res}$ &
$\sigma_{\rm LAC1}^{\rm 2-res}$ \\
\hline
CLIC & 110 & 380 & 36 & 280 & 29 & 1,200 & 74 \\
DLC & 48 & 60 & 5.5 & 49 & 4.3 & 160 & 8.3 \\
JLC & 12 & 31 & 2.8 & 23 & 2.2 & 100 & 6.1 \\
NLC & 9.7 & 21 & 1.9 & 16 & 1.5 & 67 & 4.3 \\
TESLA & 130 & 116 & 11 & 100 & 9.9 & 280 & 14 \\
\Y=0 & 2.6 & 8.4 & 0.72 & 5.6 & 0.56 & 30 & 2.5 \\
\hline
DLC & 307 & 3,000 & 360 & 1,900 & 280 & 11,000 & 1,400 \\
JLC & 24 & 250 & 30 & 150 & 23 & 950 & 140 \\
NLC & 15 & 130 & 15 & 78 & 11 & 490 & 67 \\
TESLA & 590 & 3,600 & 390 & 2,400 & 310 & 13,000 & 1,200 \\
\Y=0 & 3.7 & 24 & 2.5 & 13 & 1.9 & 91 & 14 \\
\hline
\end{tabular}
\end{center}
\vglue 0.5cm
\elevenrm
\baselineskip=14pt
\noindent
Due to the presence of numerous beam--induced backgrounds it is not clear
whether microvertex detectors can be employed at future \ee\ linacs. We
therefore borrow techniques used at hadron colliders, where the ratio of
non--charm to charm events is even larger, and require the presence of a
hard ($p_T \geq 5$ GeV), central ($|y| \leq 2$) muon in the event. We also
require the second heavy quark to be central ($|y| \leq 2$) so that its decay
can be studied, but this reduces the rate only by some 20\% once the other
two cuts have been applied. We include quark $\rightarrow$ hadron
fragmentation using the Peterson fragmentation function \cite{12} with
$\epsilon =$0.06 (0.006) for $c$ ($b$) quarks, but we use free matrix elements
to describe the decay of the quark. The resulting ``detectable'' \ccbar\ and
\bbbar\ cross sections as predicted from the DG parametrization are tabulated
in table 4 for the 500 GeV colliders. Comparison with table 2 shows that our
still rather mild cuts already reduce the direct \ccbar\ cross section by
almost a factor of 5,000 and suppress the resolved photon contribution even
more. As a result the total rate after cuts is actually dominated by the
direct contribution. Since the $p_T(\mu)$ cut increases the typical value of
Bjorken--$x$ by a factor of 10 or so, the difference between different
parametrizations of the gluon content of the photon now only leads to a factor
of 2 uncertainty for the resolved photon contribution to the \ccbar\ cross
section, or at most a 20\% ambiguity in the total rate after cuts. Finally,
comparison with table 3 shows that the \bbbar\ cross section is ``only''
reduced by a factor of 300 by our cuts. Nevertheless it remains well below the
\ccbar\ cross section, due to the smaller charge of the $b$ quark. It is
therefore questionable whether \bbbar\ production from two--photon processes
can even be identified at these colliders.
\vglue 0.5cm
\tenrm
\baselineskip=12pt
\noindent
{\bf Table 4:} Total \ccbar\ and \bbbar\ cross sections from two--photon
processes for events with a hard ($p_T \geq$ 5 GeV), central ($|y|\leq$ 2)
muon, and where the second heavy quark is also central. The resolved photon
contribution has been estimated from the DG parameterization. All cross
sections are in pb.
\begin{center}
\begin{tabular}{|c||c|c||c|c|}
\hline
Collider & $\sigma^{\rm dir}_{c \overline{c}}$ &
$\sigma_{c \overline{c}}^{\rm res}$ &
$\sigma_{b \overline{b}}^{\rm dir}$ &
$\sigma_{b \overline{b}}^{\rm res}$ \\
\hline
CLIC & 8.3 & 0.82 & 0.68 & 0.49 \\
DLC & 2.4 & 0.18 & 0.19 & 0.086 \\
JLC & 0.78 & 0.065 & 0.063 & 0.060 \\
NLC & 0.54 & 0.040 & 0.047 & 0.042 \\
TESLA & 5.2 & 0.18 & 0.45 & 0.14 \\
\hline
\end{tabular}
\end{center}
\elevenrm
\baselineskip=14pt
\vglue 0.5cm
\noindent
It has also been pointed out \cite{11} that under certain assumptions even
the total \ttbar\ cross section could be dominated by two--photon processes.
However, we showed in ref.\cite{5} that for ``realistic'' machine parameters
this is only likely to happen at very high energy, \rs $>$ 1 TeV. Here we
refine our predictions for \ttbar\ production by including 1--loop QCD
corrections. Corrections to the annihilation cross section have been known
for a long time \cite{13}, but corrections to $\gaga \rightarrow \ttbar$
have been computed only recently \cite{14}. This refinement is justified since
here the resolved photon contributions are always negligibly small, so that
there is no structure function uncertainty. In order to simplify the
calculation we have used a simple parametrization of the QCD corrections:
\be \label{e1}
\sigma_{\rm 1-loop} (\gaga \rightarrow Q \overline{Q} (g)) = \sigma_{\rm tree}
+ \frac {\alpha^2 e_Q^4} {m_Q^2} 4 \pi \alpha_s c^{(1)}_{\gamma \gamma}
(\eta), \ene
with $\eta \equiv s_{\gamma \gamma}/(4 m_Q^2) - 1$ and
\beq \label{e2}
c^{(1)}_{\gamma \gamma}(\eta) &\simeq \frac {\pi}{2} - \sqrt{\eta} \left(
\frac{5}{\pi} -\frac {\pi}{4} \right), \ \ \ \eta \leq 2.637 \nonumber \\
&\simeq 0.35 \eta^{-0.3}, \hspace*{2.cm} \eta \geq 2.637. \eeq
This parametrization describes the threshold ($\eta \rightarrow 0$) exactly,
and reproduces the full 1--loop correction to better than 10\% accuracy for
$\eta \leq 100$.
\vglue 0.5cm
\tenrm
\baselineskip=12pt
\noindent
{\bf Table 5:} Total cross sections for \ttbar\ production from
two--photon processes, for four different values of the top quark mass $m_t$.
The (small) resolved photon contribution has been estimated using the GRV
parametrization. Note that, unlike in tables 2 and 3, 1--loop QCD corrections
to the direct contribution have been included, as described in the text.
All cross sections are in fb.
\begin{center}
\begin{tabular}{|c||c|c|c|c|}
\hline
Collider & $m_t=120$ & 140 & 160 & 180 GeV \\
\hline
CLIC & 10.7 & 2.51 & 0.588 & 0.136 \\
DLC & 1.97 & 0.751 & 0.287 & 0.101 \\
JLC & 2.28 & 0.805 & 0.293 & 0.101 \\
NLC & 1.97 & 0.754 & 0.287 & 0.101 \\
TESLA & 1.98 & 0.750 & 0.287 & 0.101 \\
\Y=0 & 1.87 & 0.745 & 0.287 & 0.101 \\
\hline
DLC & 660 & 285 & 126 & 56.8 \\
JLC & 81.8 & 40.0 & 20.2 & 10.4 \\
NLC & 41.1 & 20.2 & 10.3 & 5.44 \\
TESLA & 428 & 162 & 64.3 & 26.7 \\
\Y=0 & 11.1 & 6.14 & 3.56 & 2.14 \\
\hline
\end{tabular}
\end{center}
\vspace*{5mm}
\noindent
{\bf Table 6:} Cross sections for \ttbar\ production from
\ee\ annihilation, for four different values of the top quark mass $m_t$,
including 1--loop QCD corrections. In each entry the first number is the total
cross section, and the second number the cross section after the cuts on
$m_{\ttbar}$ and on $p_{L, \ttbar}$ described in the text.
All cross sections are in fb.
\begin{center}
\begin{tabular}{|c||c|c|c|c|}
\hline
Collider & $m_t=120$ & 140 & 160 & 180 GeV \\
\hline
CLIC & 1,066 / 307 & 879 / 226 & 674 / 141 & 472 / 61 \\
DLC & 846 / 39.7 & 791 / 34.8 & 726 / 27.7 & 643 / 16.7 \\
JLC & 819 / 25.8 & 767 / 21.9 & 707 / 16.6 & 631 / 9.3 \\
NLC & 800 / 14.6 & 752 / 12.4 & 697 / 9.4 & 629 / 5.4 \\
TESLA & 897 / 90.4 & 831 / 80.2 & 753 / 64.5 & 649 / 38.9 \\
\Y=0 & 775 / 6.35 & 731 / 5.02 & 681 / 3.55 & 621 / 1.86 \\
\hline
DLC & 492 / 212 & 463 / 192 & 430 / 168 & 393 / 141 \\
JLC & 273 / 15.5 & 265 / 14.2 & 257 / 12.8 & 250 / 11.3 \\
NLC & 248 / 6.57 & 242 / 5.91 & 235 / 5.25 & 229 / 4.57 \\
TESLA & 467 / 193 & 446 / 180 & 422 / 164 & 394 / 144 \\
\Y=0 & 224 / 2.05 & 218 / 1.74 & 213 / 1.46 & 208 / 1.21 \\
\hline
\end{tabular}
\end{center}
\elevenrm
\baselineskip=14pt
\vglue 0.5cm
\noindent
The resulting \ttbar\ cross sections from two--photon and annihilation
processes are collected in tables 5 and 6, respectively. We see that for
\rs\ = 0.5 TeV the two--photon contribution is several orders of magnitude
smaller than the annihilation contribution; the same is true for the X--band
designs (JLC and NLC) even at \rs\ = 1 TeV, unless $m_t$ is near its present
lower bound. Indeed, in many cases it is far from obvious whether the
two--photon contribution can even be identified. In order to investigate this
question we have applied cuts on the annihilation contribution which are
designed to isolate events in a kinematical configuration characteristic for
two--photon processes. We first require that the invariant mass of the
hadronic system be considerably smaller than the nominal \rs\ of the collider
($<$400 GeV for \rs\ = 0.5 TeV, $<$ 600 GeV for \rs\ = 1.0 TeV). Only
annihilation events where at least one hard (beam-- or bremsstrahlung) photon
is emitted fom the initial state will pass this cut. If indeed only one
photon is emitted, the event will have a considerable longitudinal momentum.
By requiring the longitudinal momentum in the lab frame to be less
than 80 (250) GeV at \rs\ = 0.5 (1.0) TeV, we only keep events
where {\em both} the electron and the positron have emitted a hard photon. Of
course, these cuts greatly enhance the effect of beamstrahlung on the
annihilation contribution, see table 6. With the exception of CLIC all 500 GeV
colliders give more or less the same total \ttbar\ rate; however, the cross
section after cuts varies quite dramatically between different designs.
Notice in particular that the ratio of two--photon to annihilation events
after cuts is {\em largest} for the machines with {\em least} beamstrahlung.
We should mention that our cuts also reduce the two--photon cross section by
about 40\% (20\%) at \rs\ = 0.5 (1.0) TeV. At best we can therefore expect
a (two--photon) signal to (annihilation) background ratio of 1:10 (1:70) for
light (heavy) tops at \rs\ = 0.5 TeV. Since these cuts will introduce some
experimental uncertainties in the annihilation contribution, we conclude that
detection of the two--photon contribution to \ttbar\ production will at best
be marginal at this energy. On the other hand, this contribution should be
readily detectable at the 1 TeV X--band designs. Once again the significance
of the two--photon signal will be higher at low beamstrahlung machines, were
we can still expect several hundred two--photon produced \ttbar\ pairs per
year.
\vglue 0.6cm
{\elevenbf\noindent 4. Summary and Conclusions}
\vglue 0.4cm
We saw in sec.2 that two--photon production of jets can make the use of the
$Z$ peak for QCD studies and detector calibration impossible. Beamstrahlung
can lead to truely enormous charm production rates, between 0.4 and 6 billion
pairs per year already at \rs\ = 0.5 TeV even according to the rather
conservative DG parametrization. However, in all likelihood only a tiny
fraction of these events can actually be identified as being due to charm
production; the number of tagged charm events will probably not be higher than
at LEP1 and certainly smaller than at a $\tau/$charm factory. The number of
identifiable \bbbar\ events is even smaller. Finally, we saw that
beamstrahlung actually makes it {\em more} difficult to study two--photon
production of \ttbar\ pairs, since the concurrent distortion of the electron
beam spectrum makes many annihilation events look like two--photon events. This
study therefore confirms our previous conclusion \cite{5} that machines with
much beamstrahlung offer {\em no} physics advantage over designs with little
beamstrahlung.
\vglue 0.5cm
{\elevenbf \noindent 5. Acknowledgements \hfil}
\vglue 0.4cm
The work of M.D. was supported by a grant from the Deutsche
Forschungsgemeinschaft under the Heisenberg program, as well as in part by the
U.S. Department of Energy under contract No. DE-AC02-76ER00881, and in part by
the Wisconsin Research Committee with funds granted by the Wisconsin Alumni
Research Foundation.
\clearpage
{\elevenbf\noindent 6. References \hfil}
\vglue 0.4cm

\clearpage
\section*{Figure Captions}
\renewcommand{\labelenumi}{Fig.\arabic{enumi}}
\begin{enumerate}

\item  
Beamstrahlung spectra for 500 GeV (a) and 1 TeV (b) colliders. The dotted
curves show the bremsstrahlung (Weizs\"acker--Williams) contribution to the
photon flux, which only depends on the machine energy.

\vspace*{5mm}

\item  
Invariant mass soectrum of di--jets for three representative 500 GeV collider
designs. Two--photon and annihilation contributions are shown separately.

\end{enumerate}
\end{document}